\documentclass[twocolumn,pre,floats,aps,amsmath,amssymb]{revtex4}
\usepackage{graphicx,amsmath,subfigure}
\usepackage{bm}
\usepackage{natbib}
\begin{document}

\title{Construction of an inexpensive molecular Iodine spectrometer using a self developed Pohl wavemeter around 670 nm wavelength}
\author{Sachin Barthwal$^a$\footnote{corresponding author email:sachin.pranav@gmail.com}, Ashok Vudayagiri$^{a,b}$}
\affiliation{$^a$Advanced Centre for Research in High Energy Materials, University of Hyderabad \\$^b$School of Physics, University of Hyderabad, Hyderabad 500046, India}
\date{\today}

\begin{abstract}
We describe construction of an inexpensive Iodine Spectrometer with a home made Iodine vapour cell and a  self developed wavemeter based on Pohl Interferometer, around 670 nm wavelength.This can be easily realised in an undergraduate teaching laboratory to demonstrate use of a diode laser interferometry using a Pohl interferometer and measurement of wavelength using image processing techniques.Visible alternative to the IR diode lasers, 670 nm diode laser used here give chance to undergraduate students to perform comprehensive though illustrative atomic physics experiments including the Zeeman effect, the Hanle effect, Magneto Optic Rotation (MOR) effect with a little tweaking  in the present spectrometer. The advantage of the spectrometer is its ease of construction with readily available optics, electronics, evacuation and glass blowing facilities and easy analysis algorithm to evaluate the wavelength. The self developed algorithm of raster scanning and circular averaging gives the researcher insight into the basics of image processing techniques. Resolution approaching 0.5 nm can be easily achieved using such simple setup. \end{abstract} 

\maketitle
\section{Introduction} With more than 100,000 ro-vibrational lines in visible and near IR region \cite {Huang}, diatomic Iodine presents a wealth of reference lines for various spectroscopic studies making it an integral part of most of the atomic spectroscopy experiments. A large nuclear quadrupole moment and long lifetime of upper states in Iodine result in easily resolvable, though narrow transitions \cite{Borde},\cite{Cheng}. Thus it has been possible to use it for recommendation on defining meter \cite {Felder}, laser stabilisation of many systems \cite{Cornish} and other important Metrological applications. Another aspect of Iodine spectroscopy is the availability of standard atlases wherein the frequencies can be looked upon \cite{Gerstenkorn} or can be predicted by available programs like IodineSpec5 \cite{IodineSpec}   

All these facts make Iodine a suitable molecular species to be used for experimenting at the undergraduate level. It gives budding researchers a flavour of Atomic Physics experiments covering simple absorption/fluorescence spectroscopy to more interesting ones involving the Zeeman effect \cite{Goncharov}, the Hanle effect \cite{Rowley} and Magneto Optic Rotation (MOR) \cite {Budker} to name a few. As a visible laser is used for the present experiment at 670 nm, it makes the study easier in terms of detection requirement, visibly as well as electronically. Such introduction of iodine as an atomic species in undergraduate labs was not being done earlier due to the non-availability of a cheaper and simpler vapour source and usage of inexpensive diode lasers.

Here we present construction of an inexpensive Iodine spectrometer which comprises of an easy to design vapour cell along with a diode laser at 670 nm and a Pohl wave meter used for measuring the wavelength. The instrument presented here not only gives a tool to understand basic spectroscopy but also gives a flavour of diode laser usage and doing interferometry with it using a simple  image processing algorithm in MATLAB and Pohl interferometer. 

Present work is divided into 5 sections. We first introduce our preparation of Iodine vapour cell describing the nitty gritty of the design parameters involved. Next are details on describing self developed Pohl wavemeter and wavelength measurement using the same. This is followed by the Iodine spectroscopy around the 670 nm laser wherein we access transition at 671 nm. The 671 nm transition is the hyperfine components of the R(78)4-6 line of $I_2$ \cite{Huang}. We then describe the wavelength measurement of Iodine spectrometer using Pohl wavemeter. This is then followed by results and analysis and sources of error. Final section on conclusion discusses possible up-gradation  on the wavemeter design and tweaking of the Iodine spectrometer to study various atomic physics effects. We finally describe incorporating the present work in our ongoing Li laser cooling experiment.

 \section{Iodine vapour cell preparation}Iodine vapour cell was constructed using the following steps. First and foremost, all the optical components/ glasswares were thoroughly cleaned, initially with soap water and then acid and methanol. A  tube of BK7 glass, diameter 25 mm and length 70 mm was taken and its ends were fused with two optical flats, also of BK7 glass with a regularly available $\lambda/4$ flatness. Two short glass stems were fused on to the sides of this tube diametrically opposite to each other.  One of the two stems was then filled with pure iodine and sealed from outside. The other stem was coupled to a vacuum pump and evacuated to a regular rotary vacuum $\sim$10$^{-3}$ Torr($1.33\times10^{-5}$P).  Care was taken to keep the iodine stem cool at this point so that iodine is not pumped out. After this, the vacuum stem was heated to soften and then sealed. When this cell is required to be used, the iodine stem is heated so that iodine sublimes and fills the cell. Elemental Iodine being toxic, care must be taken while handling the same. A glove box was used while handling the element during the above procedure. It is noted here that as an alternative to glass blowing and evacuation facilities, one can buy a commercially available vapour cell of Iodine (for example Thorlabs GC19100-I). 

\section{POHL INTERFEROMETER: PRINCIPLE AND DETAILS OF MEASUREMENT}
\label{sec:POHL }
Pohl configuration \cite {Hecht} provides a quick and simple, though efficient way of measuring wavelength \cite {Michelson},\cite {Hariharan}. This is shown in our previous work on some preliminary measurements of wavelength of a diode laser with Pohl interferometer \cite{sachin}. Although a very simple interference device, it has been used for doing sensitive optical characterisations like a phase measurement \cite {Ewbank}. It's versatility has been proven in shop testing conditions as well while measuring parallelism of transparent surfaces \cite {Wasilik},\cite{ Malakara}. Unlike other interferometry devices \cite {Francon}, not needing very special optical components, it can very easily be incorporated in undergraduate laboratories as a wavelength measuring device.\\

\begin{figure}[ht]
\subfigure[ ]{\includegraphics[width=2 in]{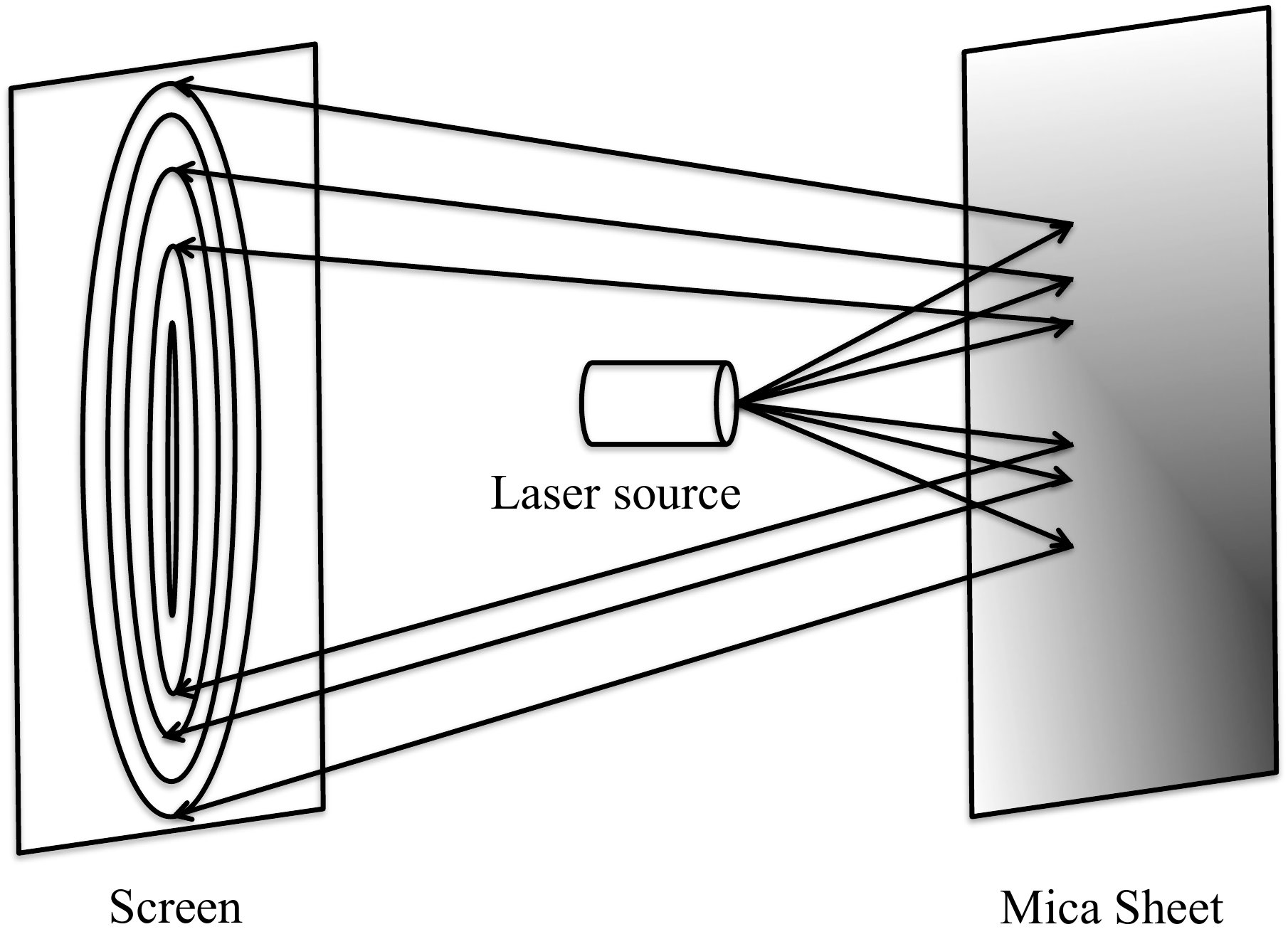}}
\subfigure[ ]{\includegraphics[width=3.5 in]{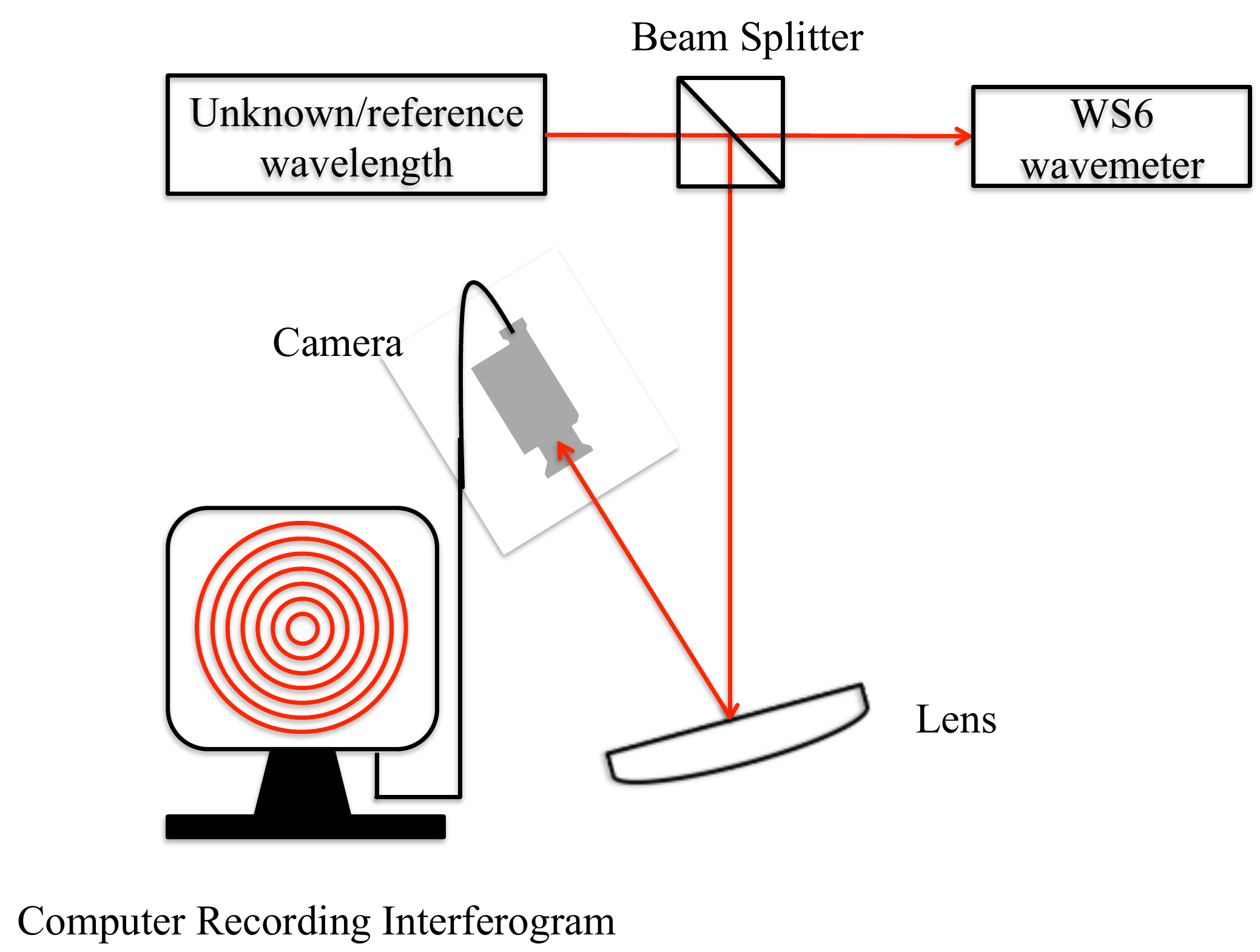}}
\caption{a:Illustration of Pohl fringes with a Mica Sheet at near normal incidence; \\
b:Experimental arrangement for Pohl interferogram measuring wavelength for unknown wavelength}
\label{fig: pohl}
\end{figure}

\begin{figure}[ht]
\subfigure[ ]{\includegraphics[width=3 in]{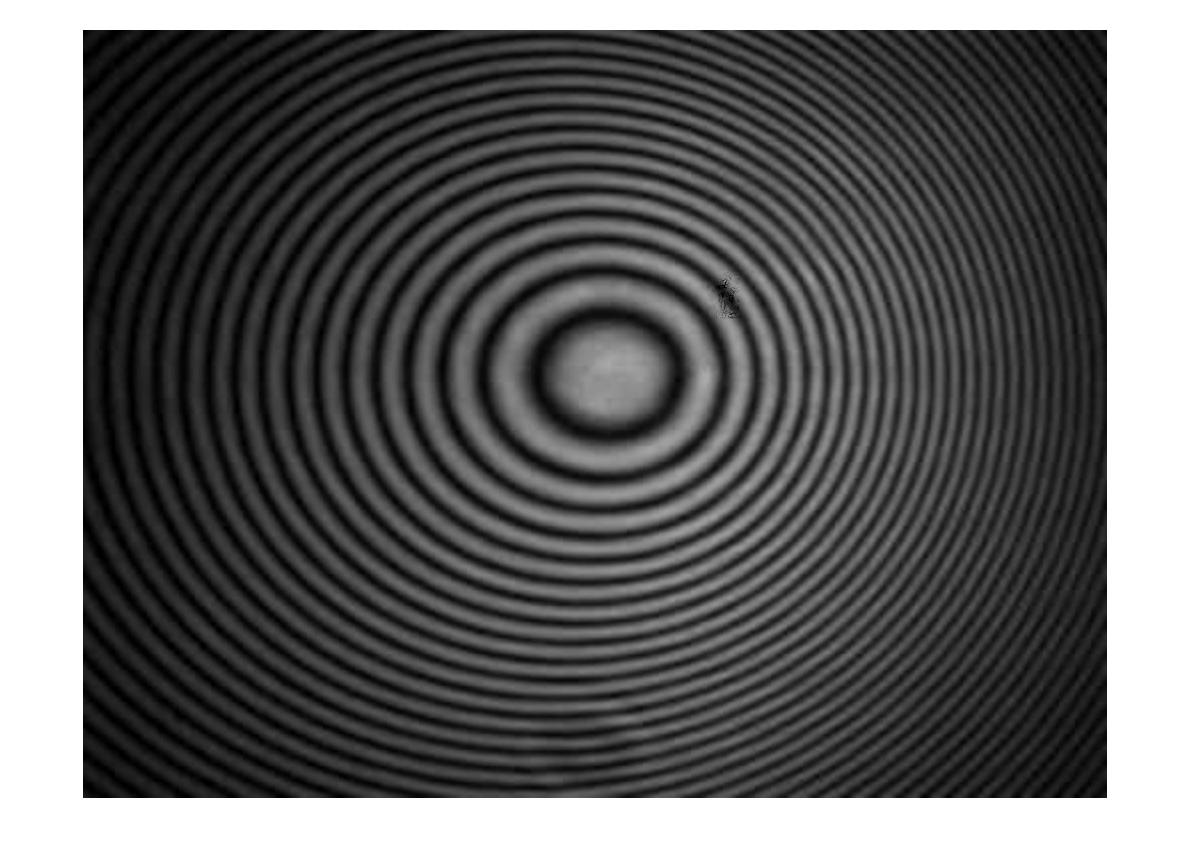}}
\subfigure[ ]{\includegraphics[width=3 in]{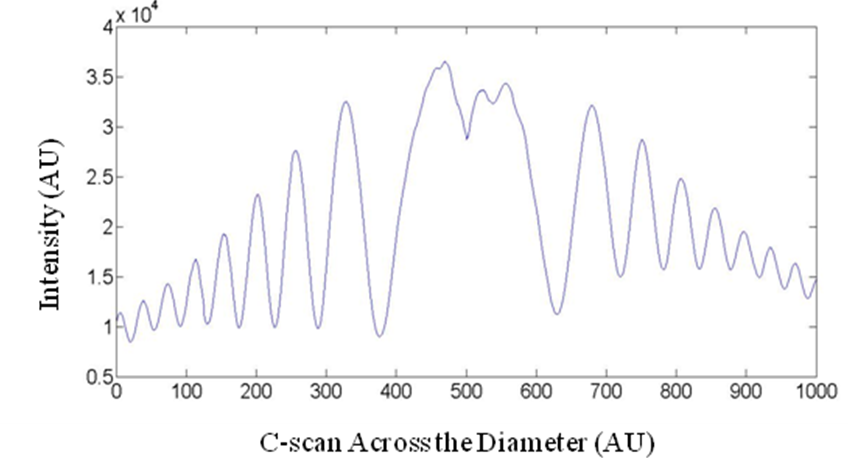}}
\caption{a:An interferogram with a 670 nm laser; b:A C-scan using the averaging algorithm}
\label{fig:ring}
\end{figure}

 Pohl interferometer shown in Fig \ref{fig: pohl}a is an amplitude splitting interferometer. As shown, a point source when reflected by front and back surfaces of a mica sheet, resembles a situation where two pseudo sources separated by double the thickness of the sheet are emitting light and result in interference pattern at a point on the screen. The full schematic showing wavelength measurement using Pohl wavemeter is in Fig. \ref{fig: pohl}b. A plano-convex lens of $f$ $\sim$1000 mm is taken for producing Pohl fringes. A DCU223M Thorlabs Digital Camera with 30 frames per second is used to capture the images. At the heart of the wavemeter lies post detection analysis of the interferogram. This involves a simple but efficient self developed algorithm of raster scanning and then circular averaging. Raster scan is a common place technique for surface studies using AFM in the field of soft condensed matter physics \cite{Chang}. In the algorithm, we first choose region of interest (ROI) as first few fringes (say seven) going outward in the interferogram and it's centroid ($x_c$,$y_c$) is found. Then we take a radial point ($x_r$,$y_r$) on the outermost fringe in the ROI. Joining ($x_c$,$y_c$) and ($x_r$,$y_r$), we then find intensity profile along this radius thus getting a measure of radial variation of the intensity in the rings formed. The same procedure is followed to give similar data along radius with different inclinations with respect to the first one. This is done by adding an extra variable $\theta$, an angle between the initial radius and the next one. This procedure is followed for $\theta$ going from 0 $\to$$360^o$ for every 5 degrees angle and all intensity variations are found. This is finally averaged to get a circularly averaged intensity profile. One such interferogram and intensity variation found experimentally is as shown in Fig \ref{fig:ring}. Whole of this algorithm is realised using MATLAB image processing package. 
                                                                                                                                                                          
 First the above procedure is followed with the He-Ne laser at 632 nm. Now by using the interferometer Ring formula \cite{Ghatak} and knowing the wavelength of the light creating the interferogram we can measure the other constant i.e. $k$ in the equation below

\begin{equation}
\lambda= k \frac{D_{m+p}^2-D_m^2}{4p}
\end{equation}

where $D_{m+p}$ and $D_m$ are diameters of $(m+p)^{th}$ and $m^{th}$ fringe in pixels while $k$ is only the calibration factor. Similar $k$ value is measured for the other He-Ne laser at 543 nm to take care of any dispersion effects. This $k$ value is then used for the measurement of unknown wavelength for the diode laser. As an alternative to 543 nm He-Ne, one can use a simple and inexpensive laser pointer at 532 nm which is frequency doubled output from a DPSS laser.

\section{Iodine spectroscopy at 670 nm transition using Pohl wave meter}
\label{sec:I spectroscopy}

\begin{figure}[ht]
\subfigure[ ]{\includegraphics[width=3 in]{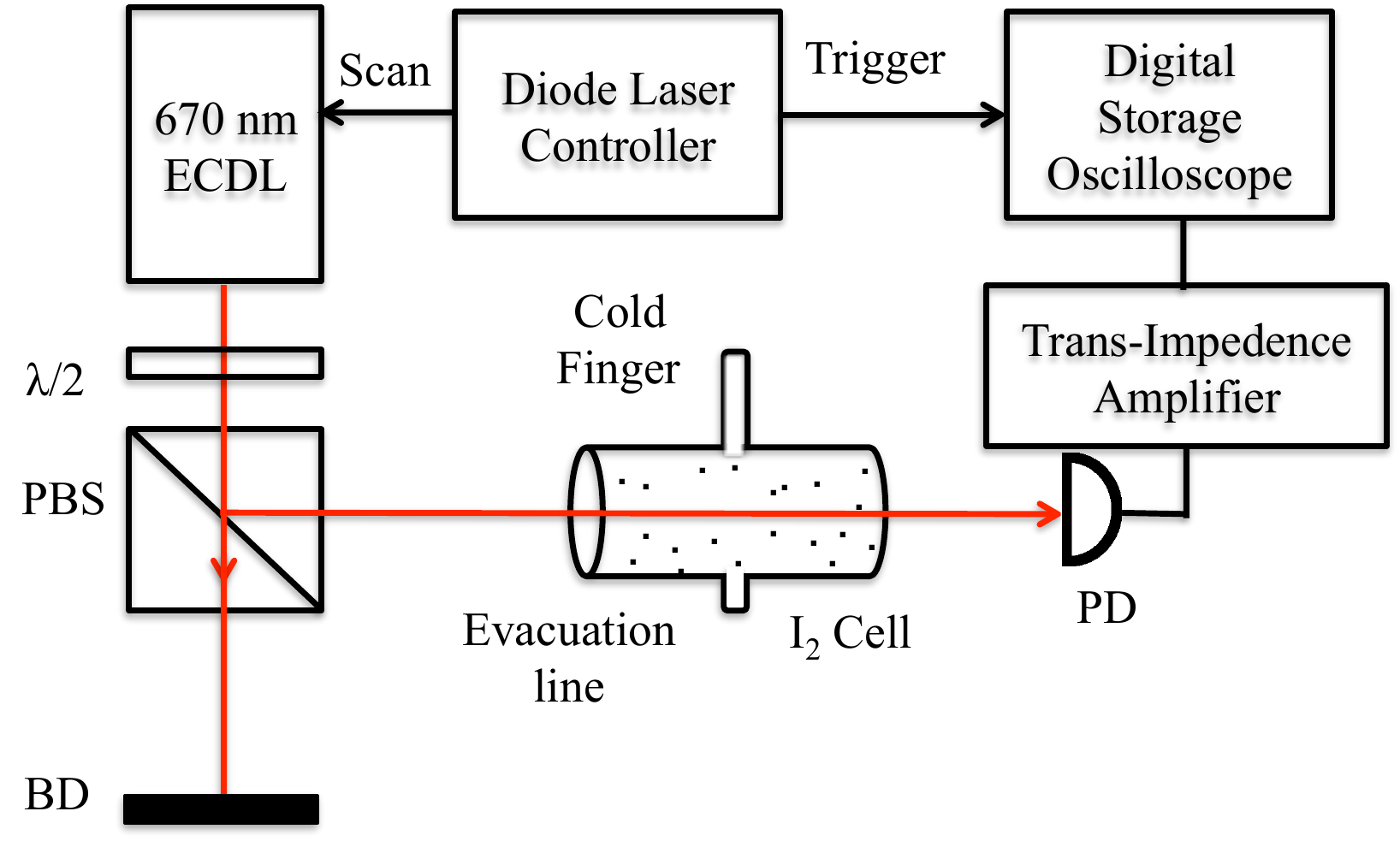}}
\subfigure[ ]{\includegraphics[width=3 in]{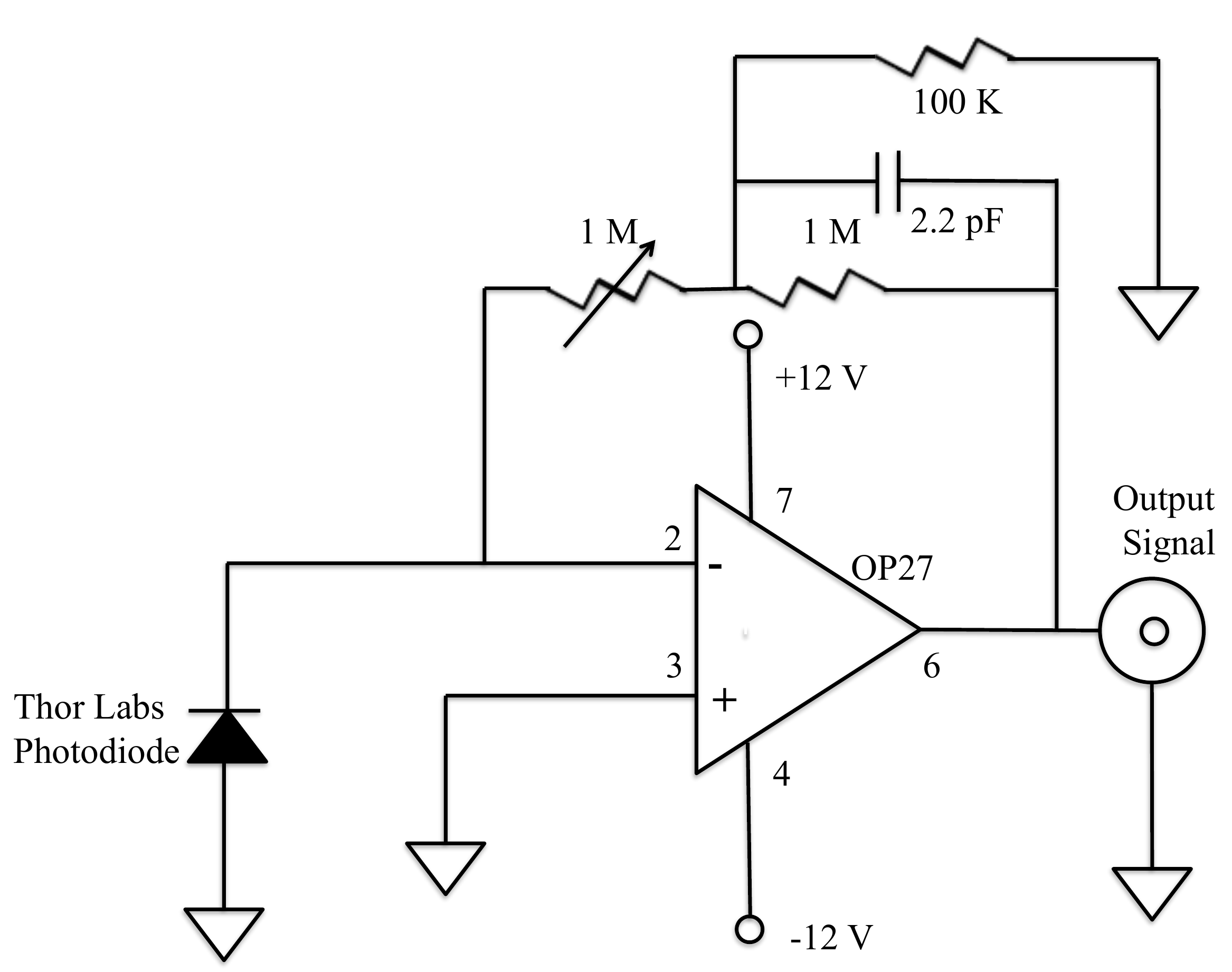}}
\caption{a:Experimental arrangement for Iodine Spectroscopy; PBS: Polarising beam splitter, BD: Beam Dump, PD: Photodiode, $\lambda/2$:Half Waveplate; \\
b:High gain Trans-impedence amplifier used for signal detection }
\label{fig: setup}
\end{figure}

Laser used for the experiment is a 35 mW external cavity diode laser (ECDL) by Toptica Photonics centred at 670 nm with tunability of $\sim$15 nm around the centre. This offers opportunity to access various lines in the Iodine molecule by varying the diode frequency utilising the tunability of the laser. The experimental arrangement is shown in Fig.\ref{fig: setup}a.  As an alternative to the commercial ECDL researchers can follow many self developed designs of ECDL and related electronics, refer for example \cite{wieman}. 

The ECDL output is sent through the home made iodine vapour cell described in the previous section. For tuning the diode laser, we needed to play with the grating angle initially using an adjustment screw behind the grating plate. It pushes the grating to change the incident angle of the laser beam on the grating. For finer adjustment of the frequency, piezo drive going to the piezo stack behind the grating plate is played with in addition to the diode laser current. Once the diode is tuned to the right frequency one can easily see the fluorescence with naked eyes. This can be made brighter by heating the cell a little using a heat gun to increase the Iodine vapours in the cell. However excessive heating is avoided so as not to coat the optical windows with violet vapours of the molecule. The cell is heated to about 150$^0$C using heating tapes tied around to get good enough vapours.

To get a good S/N ratio one needs to play with power of the input beam by rotating the half wave plate. The input power ($P_{in}$) has to be lowered to a level so that the relative absorption of the beam with respect to the $P_{in}$ becomes sufficient to get good enough absorption peak heights. Another thing to take care is the mode hops of the laser. We find that the laser could be scanned by around 2 GHz without any mode hops problem. For recording the absorption spectra we have used a 1 nsec rise time photodiode (DET10A/M Si- Biased detector from www.thorlabs.com) available in our laboratory. Alternatively an inexpensive photodiode BPW34 can as well be used which is sufficient to fulfil the demand of such experiment. This is then connected to a trans-impedence amplifier built around an easily available low noise precision op-amp OP27 as shown in Fig. \ref{fig: setup}b. It is to be noted in the design of the transimpedence amplifier, a constant gain-bandwidth product would give a limit to the feedback resistor value being used. An unusually high gain resistor might give a bandwidth which is out of our region of interest and the frequency response will go bad. To compensate for this we have used OP27, a wide bandwidth op-amp. In addition to this we used a feedback capacitor to reduce the gain for noise that is outside the bandwidth of the absorption signal \cite{Horowitz}. Here it should be noted that for increasing bandwidth of the detection system, an opamp faster than OP27 and a lower trans-impedence gain will help. Further a second stage voltage amplifier can be added to get good signal at the oscilloscope.  

\begin{figure}[ht]
\includegraphics[width=3.5 in]{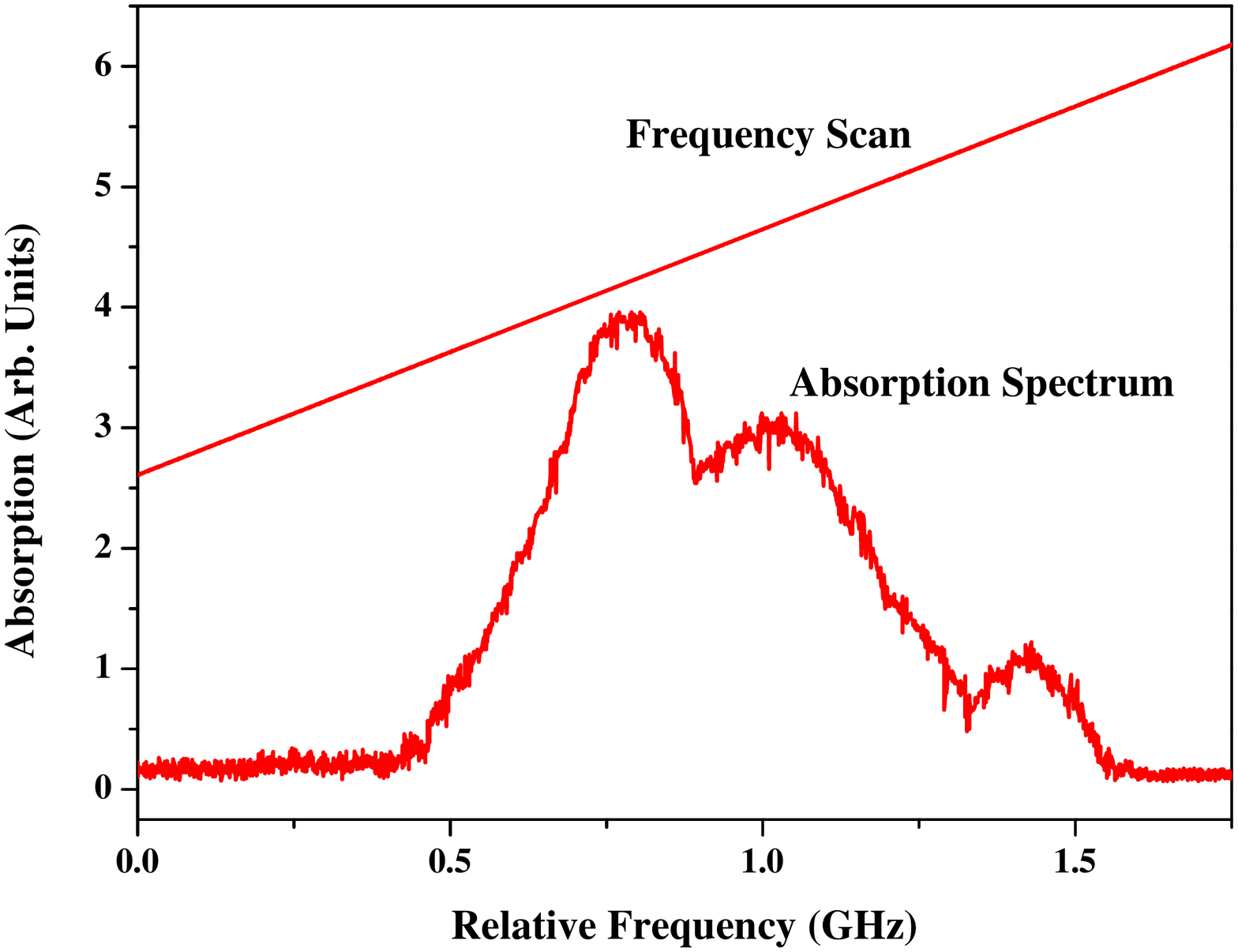}
\caption{Absorption spectrum with laser scanning around 670 nm (Piezo scan shown on top of it) }
\label{fig: spec}
\end{figure}   
A typical absorption signal for Iodine from the current setup with the Thorlabs photodiode  is as shown in Fig \ref{fig: spec}. As seen, the laser is scanned by $\sim$ 1 GHz which is well within the mode hop free region of the diode. The lines are identified as $a_1$-$a_{15}$ hyperfine components of the R(78) 4-6 line of $^{127}I_2$ using IodineSpec5 \cite {IodineSpec}. The lines get broadened  mainly due to the Doppler broadening at the temperature of 150 $^0$C and pressure broadening due to high vapour density. At 150 $^0$C the Doppler broadening comes to about $\sim$0.6 GHz using the below equation \cite{Demtroder}. 

\begin{equation}
\delta\nu_D=7.16\times10^{-7}\nu_0 \sqrt{\frac{T}{M}}
\end{equation} 
where T is the temperature in K and M is the Molar mass.    

Heating the cell at 150 $^0$C results in a vapour pressure of around  $\sim$5 P using equation \cite{Gillespie}

\begin{equation}
log(P) = \frac{-3512.830}{T}-2.013 log(T)+ 18.37971
\end{equation}
where T is the temperature of the cell in K and P is the resulting vapour pressure in P. This gives a broadening of around $\sim$50 KHz \cite{Huang}. As Doppler's broadening is  $\sim$0.6 GHz, it masks most of the spectrum and the final outcome shows up as three peaks only as seen in the spectrum.

  \section{Results, Analysis and Sources of Error}
\label{sec:Results, Analysis and Sources of Error}

Measurement results are summarised in Table1. We measured the wavelength of all the three peaks resolved by the spectrometer. As the peaks are just $\sim$ 100 MHz away from each other, it is not in our resolution limit to resolve the same using the Pohl interferometer measurement and hence each peak measurement shows the same wavelength. Each point is a result of 10 successive data sets taken to avoid any error in finding the diameter of the interference ring. As a consistency check we have measured wavelength using different pairs of rings (see Table1).  We have chosen $1^{st}$ and $5^{th}$ and $3^{rd}$ and $7^{th}$ rings as the visibility of the rings will diminish as we go outward from centre. This is due to the fact that rings are getting formed by multiple front and back surface reflections. This results in the intensity difference between the two reflections becoming more and more prominent as we go radially outward. This goes on till the time fringe contrast goes very low and there is uniform illumination with no more fringes seen on the screen. We have further compared the measurement with a commercial wavelength meter WS6 Angstrom wavemeter \cite{Wavemeter} which is quoted to be with relative accuracy of $10^{-6}$. Our values come within the resolution we claim here.  

\begin{table}
\caption{Wavelength measurement using the Pohl wavemeter}
\begin{ruledtabular}
\begin{flushleft}
\resizebox{9 cm}{1.5 cm}{
\begin{tabular}{cccc}
Peak No. &  WS-6 Value & \multicolumn{2}{c}{Pohl Interferometer value} \\ \hline
 && \scriptsize {With $1^{st}$ and $5^{th}$ ring}& \scriptsize{With $3^{rd}$  and $7^{th}$  ring}\\ \hline
1 & 670.9 & 670.8 & 671\\
2 & 670.9 & 670.6 & 670.9\\
3 & 670.9 & 671 & 670.8 \\
\end{tabular}}
\end{flushleft}
\end{ruledtabular}
\end{table}

Lower pixel resolution of the CCD used (pixel size $\to$ 4.65 $\mu$m x 4.65 $\mu$m) becomes the main source of error. Replacement of the same with a higher resolution CCD camera is to be taken in the next version of the instrument to reduce this error. Exact measurement of $D_m$ in eq.1 gives another possible source of error. This is tried to be reduced by taking data multiple times (10 times in the experiment) and making an average. To avoid any error coming due to visibility of fringes, different set of interference rings for wavelength measurement are used(see Table 1). Dispersion error can come due to difference between wavelength of calibration lasers and actually measured wavelength. Instability of the gas laser used, due to fluctuation in the output can be another source of error. However all these errors come within the resolution we claimed here i.e. 0.5 nm \cite {Elden}, \cite {Siegman}.

  \section{CONCLUSION}
\label{sec:CONCLUSION}

   In conclusion we can say that we have presented construction of an Iodine spectrometer. A simple method for making an Iodine vapour cell was presented. This can very easily be repeated in a University workshop. Transition R(78) 4-6 is then accessed in $^{127}I_2$ around 670 nm. We then measured the transition wavelength using an in-house, simple, portable and accurate wavemeter, using a self developed image processing technique and Pohl configuration as a simple Interferometer. We can further better the accuracy of the method using a fibre coupled beam so as to get a refined wavefront and hence a clearer and sharper interferogram. We further plan to have a high resolution CMOS camera to be used for detection part to enhance the image quality. Finally we are planning to automate the whole process of measurement using a lab view code with a GUI as well to make a module for wavemeter measurement.  With a little tweaking,  iodine spectrometer designed here can be used to perform atomic physics experiments including the Zeeman effect, the Hanle effect and MOR. This can be easily achieved by having a simple wire wound solenoid around the cell to have required magnetic field. Alternatively we can have Helmholtz configuration coils, regularly used in laser cooling experiments to achieve uniform magnetic field for such experiments.  As a next step to our Iodine experiment, we would like to incorporate the Iodine spectrometer with our ongoing Lithium laser cooling experiment. As the 671 nm Iodine transition is just 3 GHz away from the $D_2$ cooling line in Lithium \cite {Huang},\cite{Das} we can get a reference for the cooling line by measuring the beat frequency of the two lasers with a fast photodiode.
\section{ACKNOWLEDGEMENT}
\label{sec:ACKNOWLEDGEMENT} 
   We thank Mr. Gautam Karve for helpful discussions during preparation of the manuscript. The Board of Research for Fusion Science \& Technology, Nuclear Fusion Program, India is gratefully acknowledged for the financial support. Horst Knoeckel is acknowledged for help with the program IodineSpec5.  Further, model numbers mentioned in the text are for examples only and similar products from alternative sources can very well be used.

\end{document}